\documentstyle[prl,aps,multicol,epsfig]{revtex}

\begin{document}

\draft
\title{\lq\lq Barber pole turbulence\rq\rq in large aspect ratio
Taylor-Couette flow} 
\author{A. Prigent and O. Dauchot}
\address{Service de Physique de l'Etat Condens\'e, CEA Saclay, 91191 Gif
sur Yvette, France}
\date{\today}
\maketitle

\begin{abstract}
Investigations  of counter-rotating Taylor-Couette  flow (TCF)  in the
narrow  gap  limit  are  conducted   in  a  very  large  aspect  ratio
apparatus.  The  phase  diagram  is  presented and  compared  to  that
obtained  by   Andereck  {\it  et   al.}~\cite{Andereck}.  The  spiral
turbulence  regime is studied  by varying  both internal  and external
Reynolds numbers. Spiral turbulence is  shown to emerge from the fully
turbulent regime via a continuous transition appearing first as a {\it
modulated} turbulent  state, which  eventually relaxes locally  to the
laminar  flow. The  connection with  the intermittent  regimes  of the
plane Couette flow (pCf) is discussed.
\end{abstract}

\pacs{ {47.20.Ft} 
       {47.20.Ky} 
       {47.54.+r} 
       \and {47.27.Cn} 
}

\begin{multicols}{2} 
\narrowtext
The  \lq\lq  barber  pole  structure  of  turbulence\rq\rq~\cite{Feyn}
between two counter-rotating cylinders, also called spiral turbulence,
is commonly  described as alternating  helical stripes of  laminar and
turbulent flow.   There are few quantitative studies  of this puzzling
regime, where  long range order coexists with  small scale turbulence.
In  early  studies  Coles and  Van  Atta~\cite{Coles,VanAtta,Coles-VA}
measured  the  spiral shape  in  the  mid-plane  perpendicular to  the
cylinder axis  and the pattern  rotation rate. Later Andereck  {\it et
al.}     reported     qualitative     observations    when     varying
$R_i$~\cite{Andereck}.    Hegseth    {\it   et   al.}~\cite{Heg1,Heg2}
described   spiral   turbulence   within   the  framework   of   phase
dynamics.  All these studies  were limited  by their  relatively small
size.  Only one  helical turbulent stripe, winding no  more than twice
along the cylinder axis, could  be observed. Altogether, the origin of
this flow pattern remains unknown.

Performing measurements in large  aspect ratio Taylor-Couette flow, we
show that the spiral turbulence bifurcates continuously 
from  the turbulent  flow, appearing  as  a modulated  turbulent
state.  After a  rapid  description  of the  experimental  set up,  we
present  the phase  diagram  and compare  it  to the  one obtained  by
Andereck  {\it  et   al.}\cite{Andereck}  with  a  different
cylinder radius ratio.  Then  we describe the successive steps leading
to the  fully turbulent flow before discussing  the origin  of spiral
turbulence. Finally,  we examine its breakdown  into a spatio-temporal
disordered  regime   similar  to  the   laminar-turbulent  coexistence
dynamics observed in plane Couette flow~\cite{DauDav,BottChat}. 

Let  us  recall  here  the  characteristics of  the  experimental  set
up~\cite{TCvisu}. Our  TC apparatus,  two coaxial cylinders  of radius
$r_i=49.09 \pm  0.005$ mm and $r_o=49.95  \pm 0.005$ mm  with a useful
length  $L=380 \pm  0.1$ mm,  rotating independently,  has a  gap size
$d=0.863  \pm  0.01$ mm,  large  aspect  ratio $\Gamma_z=L/d=442$  and
$\Gamma_{\theta}=\pi(r_i+r_o)/d=362$, and  a radius ratio $\eta=0.983$
very close  to 1.  In this paper,  length and  time units are  $d$ and
$d^2/\nu$.  $\eta$ being fixed, the  flow is governed by the inner and
outer   Reynolds   numbers  $R_{i,o}=r_{i,o}\Omega_{i,o}d/\nu$,   with
${\Omega_{i,o}}$  the angular  velocities, and  ${\nu}$  the kinematic
viscosity.  The accuracy of $R_{i,o}$ is better than 3$\%$. 

We  visualize  the  flow  by  a  \lq\lq  fluorescent  lighting  \rq\rq
technique~\cite{TCvisu} developed  for this study.  The  water flow is
seeded with  Kalliroscope AQ 1000 ($6 \times
30 \times 0.07  \mu m$ platelets). The inner cylinder  is covered by a
fluorescent  film  and  the   entire  apparatus  is  UV-lighted.   The
fluorescent  film  re-emits a  uniform  visible lighting,  transmitted
through the fluid layer: the  more turbulent the flow, the brighter it
appears. As  the gap is  very thin, the Kalliroscope  concentration is
increased up to $25\%$ by volume to enhance the contrast. A 
rheological study has shown that  the fluid remains Newtonian, so that
the  only impact  is  a viscosity  increase  up to  $\nu=1.13~10^{-6}$
m$^2$/s at $20^\circ$C.  The  flow is thermalized by water circulation
inside the  inner cylinder. At thermal equilibrium 
the temperature  is uniform in space  up to $0.1^\circ$C  and does not
vary more than $0.1^\circ$C/hour. Images and spatio-temporal diagrams
(temporal recording of one line along the cylinder
axis) are recorded by a CCD  camera. Two plane mirrors reflect the two
thirds of the flow hidden to  the camera so that the whole cylindrical
flow can be reconstructed.

\begin{figure}
\vspace{-1.cm}
\begin{center}
	\hbox{
	\hspace{-0.35 cm}
	\psfig{figure=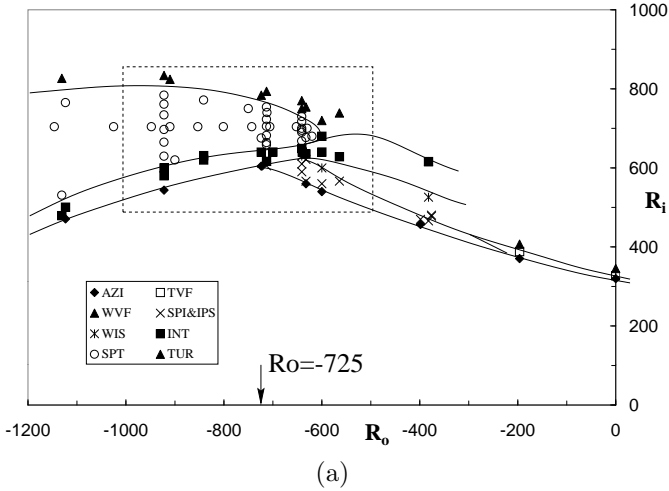, width=9cm, angle=0}
	}
        \vspace{-0.2cm}
	(a)
	\hbox{
	\hspace{-0.30 cm}
	\psfig{figure=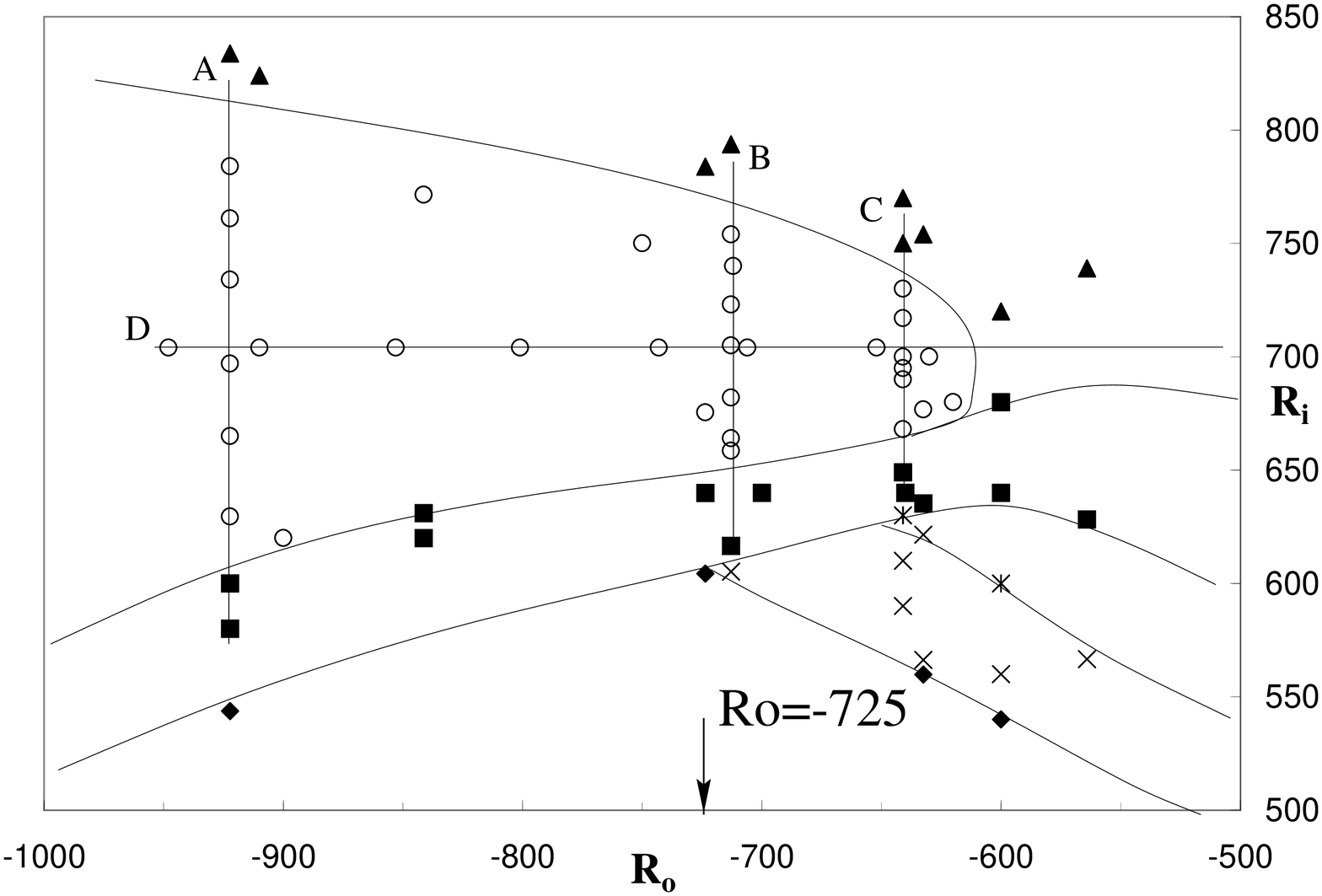, width=9cm, angle=0}
	}
        \vspace{-0.15cm}
	(b)
\end{center}
\vspace{-0.2cm}
\caption{Experimental  phase diagram  ($\eta=0.983$).  (b) displays  a
zoom  of   the  dotted  rectangle   of  (a).  The  labels   stand  for
AZI~:~azimuthal flow; TVF:~Taylor  vortex flow; WVF:~wavy vortex flow;
SPI\&IPS:~spiral  and   interpenetrating  spiral  vortices;  WIS:~wavy
interpenetrating  spiral   vortices;  INT:~intermittency;  SPT:~spiral
turbulence; TUR:~turbulence. The arrow indicates the $R_o$ value below
which the AZI  flow jumps directly to INT (see  text for details). The
solid straight lines  in (b), show the paths  along which measurements
are    conducted    (A:~$R_o=-922$;   B:~$R_o=-713$;    C:~$R_o=-641$;
D:~$R_i=704$).}
\label{fig:phasediag}
\end{figure}

Coles~\cite{Coles}  has  stressed   the  variety  of  flow  structures
observed  according to  the  path followed  in  parameter space.   The
experimental  procedure followed here  is the  most frequent  one. The
outer cylinder  is slowly accelerated to the  desired angular velocity
while the  inner one is  at rest. Then,  the inner cylinder  is slowly
accelerated  ($(1/\Omega_i).(\partial{\Omega_i}/{\partial  t})=0.03\%$
per  second) in  order  to ensure  quasi-static  transitions. (It  was
checked  that the  Taylor  vortices threshold  is  determined with  an
accuracy better  than $1\%$ as soon  as the accelaration  rate is less
than $0.05\%$).  Recording the various flow regimes in parameter space
$(R_i    ,R_o    )$,   gives    the    experimental   phase    diagram
(fig.~\ref{fig:phasediag}(a)) to be compared to that given by Andereck
{\it  et al.}~\cite{Andereck} for  a different  geometry ($\eta=0.878$
and much  smaller aspect  ratio).  The same  regimes are  observed but
their locations in the parameter space change.  The linear thresholds
increase (at $R_o=0$, the Taylor vortices appear for $R^*_i=320$ instead of
$R^*_i=120$). This was expected in regards of the
plane Couette flow limit and agrees very well with Snyder's law
$R^*_i~=~27[(1-(\eta))\eta]^(-5/3)$~\cite{Snyder}. 

\noindent
Accordingly, the other vortex flows  (WVF, SPI, IPS...) also appear at
higher  $R_i$  values.    Second,  the  laminar-turbulent  coexistence
regions (INT  and SPT),  extend to higher  $R_o$ values.  Finally, for
$R_o<-725$,  the azimuthal  flow jumps  directly to  INT  whereas, for
$R_o>-725$, it bifurcates first to IPS or SPI.

\begin{figure}
\vspace{-0.5cm}
  \begin{center}
  \hbox{
    (d)
    \psfig{figure=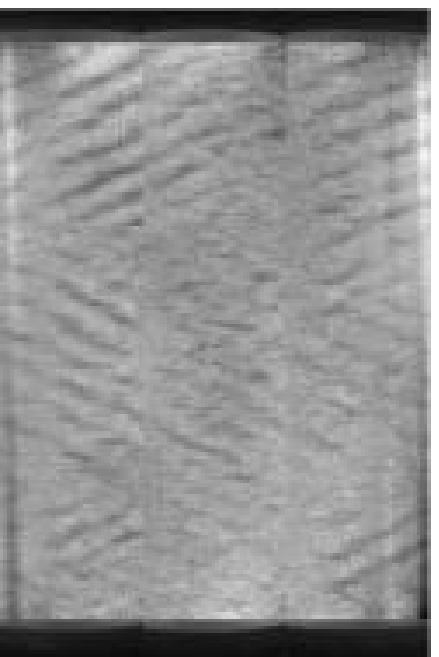, height=5.2cm, width=3.47cm, angle=0} 
    (h)
    \psfig{figure=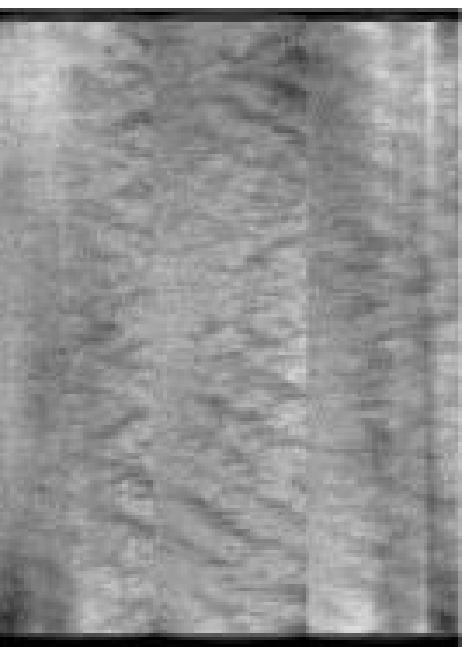, height=5.2cm, width=3.47cm, angle=0}
   }    
  \hbox{
    (c) 
    \psfig{figure=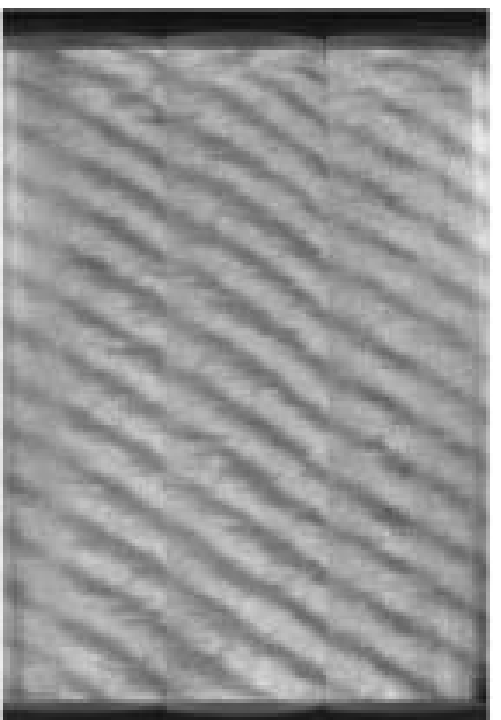, height=5.2cm, width=3.5cm, angle=0} 
    (g)
    \psfig{figure=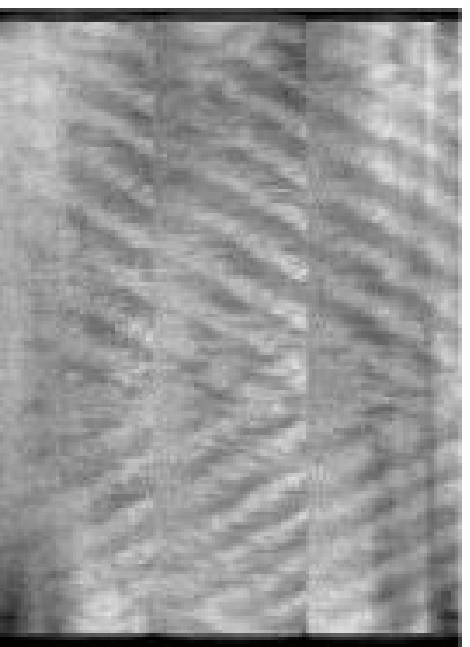, height=5.2cm, width=3.5cm, angle=0}
   }
  \hbox{
    (b) 
    \psfig{figure=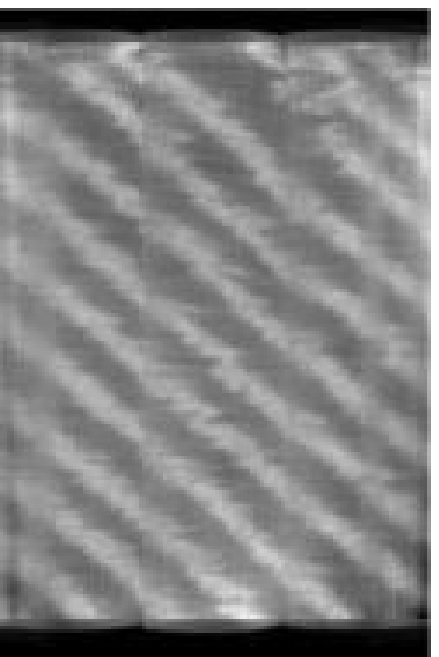, height=5.2cm, width=3.5cm, angle=0} 
    (f)
    \psfig{figure=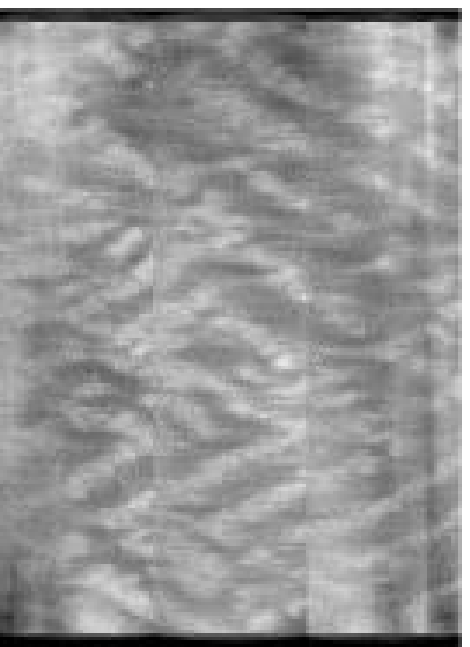, height=5.2cm, width=3.5cm, angle=0}
   }
  \hbox{
    (a) 
    \psfig{figure=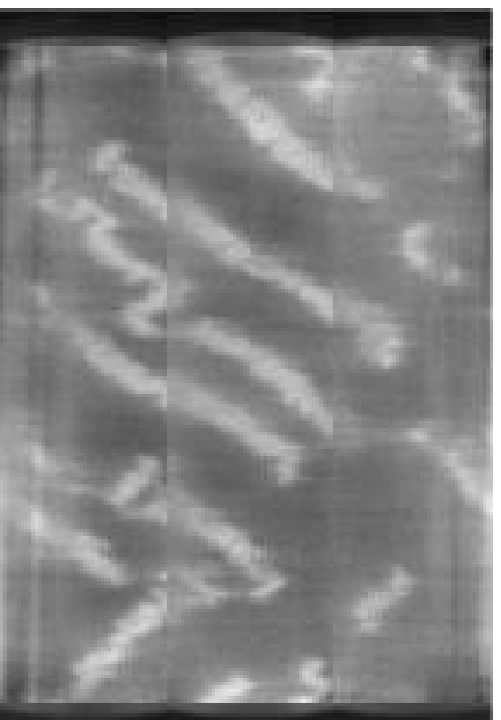, height=5.2cm, width=3.5cm, angle=0} 
    (e)
    \psfig{figure=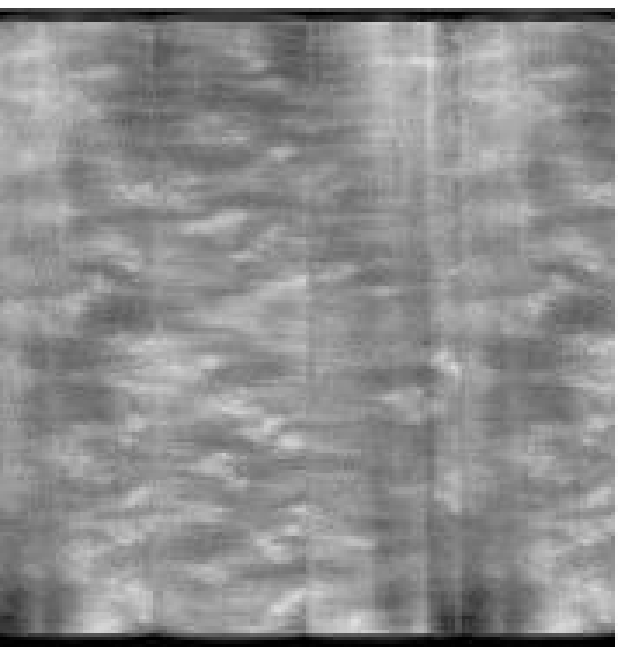, height=5.2cm, width=3.5cm, angle=0}
   }
\end{center}
\vspace{-0.5cm}
    \caption{Turbulent spots and stripes along paths A and C. 
Left column is path A, $R_o=-922$:~$R_i=$~(a)~580, (b)~600, (c)~697, (d)~784.
Right column is path C, $R_o=-641$:~$R_i=$~(e)~649, (f)~668, (g)~695, (h)~717.
Each picture displays a 360 $^\circ $ view of the whole flow ($38$ cm
high and $31.4$ cm wide.} 
    \label{fig:deroule}
\end{figure}

Figure~\ref{fig:deroule},  shows snapshots  of the  whole  flow, along
paths  A  and C.  For  $R_o=-922$,  the  boundaries trigger  turbulent
stripes  are  at $R_i$  lower  than  the  expected linear  instability
threshold  (fig.~\ref{fig:deroule}(a)).  In  fig.~\ref{fig:deroule}(b)
periodically   organized  turbulent   stripes   have  formed:   spiral
turbulence is  established as a regular rotating  pattern with uniform
pitch angle along the cylinder axis.

\begin{figure}
\vspace{-0.8cm}
  \begin{center}
    \psfig{figure=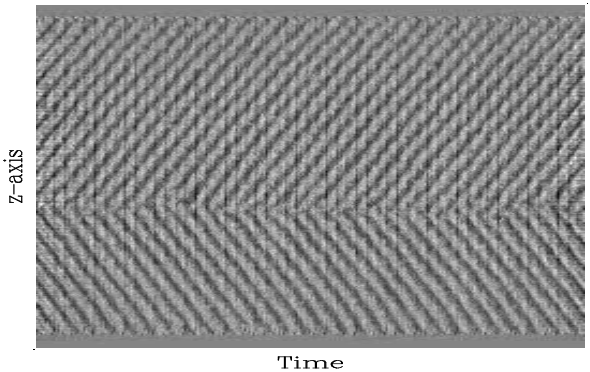, height=5cm, width=8cm, angle=0}
  \end{center}
\vspace{-0.1cm}
    \caption{Spatio-temporal diagram of spiral turbulence
displaying a front between two domains of opposite pitch
angle. $R_o=-838$, $R_i=707$. The z-axis display the whole flow height
($38$ cm) and the total recording length is $20.48$ s.}
    \label{fig:dst}
\vspace{-0.2cm}
\end{figure}

\noindent
Further increasing $R_i$, the pattern maintains  itself and the
number of stripes  increases (fig.~\ref{fig:deroule}(c)). Finally, the
pattern disappears  smoothly as homogeneous  turbulence takes over
(fig.~\ref{fig:deroule}(d)). In this regime, one notices several domains of  opposite pitch  spirals.
For  $R_o=-641$, the azimuthal flow first bifurcates to a vortex flow,
within which, further increasing $R_i$,  turbulent spots follow an
intermittent bursting dynamics (fig.~\ref{fig:deroule}(e)).   In
fig.~\ref{fig:deroule}(f), the spots grow, merge and
build  up  the turbulent  spiral  pattern.   For 
illustration,  we  show,  in fig.~\ref{fig:deroule}(g),  two  opposite
pitch  spirals competing and  leading to a so-called ``V''-shaped
pattern.   In figure~\ref{fig:deroule}(h), spiral  turbulence turns
into  uniform   turbulence  as  for   $R_o=-922$.   Fig.~\ref{fig:dst}
displays the spatio-temporal dynamics  of a front between two opposite
spirals.  This front, moving along the cylinder axis, is eventually
absorbed at one of the extremities. 

Let  us now  consider  the previous  scenario  in reverse.  Decreasing
$R_i$,  spiral  turbulence appears  from  homogeneous turbulence.   At
$R_i=R_{ic}$,  a low amplitude  modulation, with  wave vector  $k$ and
frequency $\omega$, uniformly emerges  from the turbulent flow with no
prefered direction.   A slow front dynamics occurs  between domains of
opposite  pitch  until the  $z  \rightarrow  -z$  symmetry breaks  and
selects  a  single  uniform  spiral.  Further  decreasing  $R_i$,  the
modulation   amplitude  increases  until   a  more   complex  dynamics
occurs.  The  light  intensity,   which  in  the  first  approximation
corresponds to  the turbulence  level, is recorded  in spatio-temporal
diagrams such  as fig.~\ref{fig:dst}. In the following,  we obtain the
amplitude $A$,  the axial component $k_z$  of the wave  vector and the
frequency   $\omega$   of    the   light   intensity   modulation   by
demodulation~\cite{Croq}.   Owing to  the  azimuthal periodicity,  the
azimuthal         component         of         the         wavevector,
$k_\theta=2.n_\theta/(r_i+r_o)$, is  quantized, $n_{\theta}$ being the
number of pattern periods along one perimeter.

Four  paths  in the  $(R_o,  R_i)$  plan  are investigated.  We  first
consider the three  vertical paths A, B and  C ($R_o=-922$, $R_o=-713$
and $R_o=-641$).

\begin{figure}
\vspace{-0.3cm}
  \begin{center}
  \hbox{
    \psfig{figure=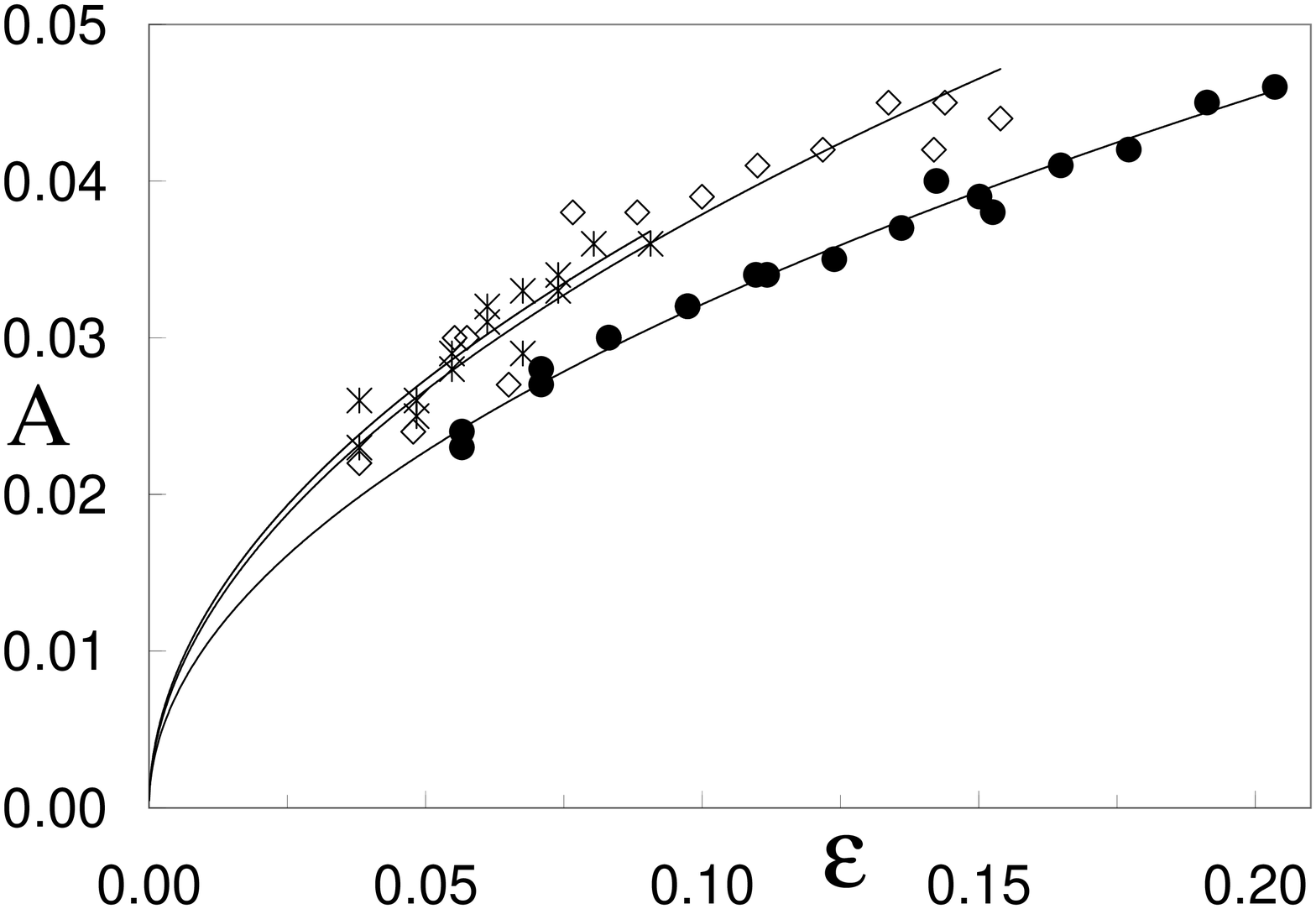, height=4cm, width=4.25cm, angle=0}
    \psfig{figure=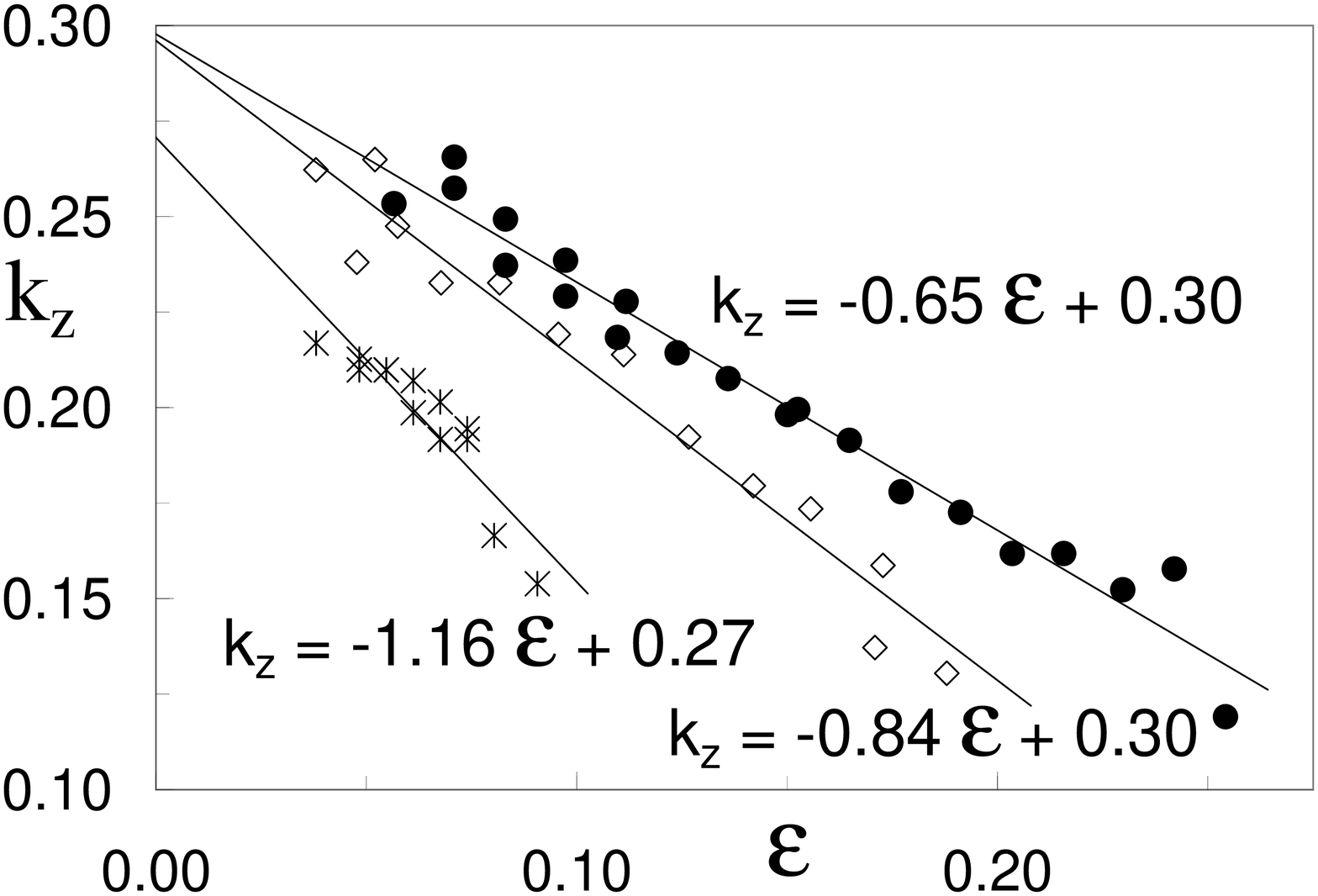, height=4cm, width=4.25cm, angle=0}
       }    
\vspace{-0.1cm}
  \hbox{
    \hspace{1.77cm} (a) \hspace{3.77cm} (b)	
       }
\end{center}
\vspace{-0.5cm}
    \caption{The spiral turbulence amplitude (a) and vertical wave
number (b) vs. $\epsilon$. ($R_o=-641$ ($\ast$), $R_o=-713$
($\diamond$), $R_o=-900$ ($\bullet$)).} 
    \label{fig:curves}
\end{figure}

\noindent
Fig.~\ref{fig:curves}(a) displays  the amplitude of the wave against
the threshold distance $\epsilon= \left|R_i-R_{ic}\right|/R_{ic}$.
$R^c_i$ is determined by extrapolation, assuming a
linear dependence of $A^2$  on  $R_i$,  as  suggested  by  the
general  framework  of amplitude equations. The resulting values,
$R_{ic}=820$ for $R_o=-922$, $R_{ic}=785$ for  $R_o=-713$ and
$R_{ic}=740$ for $R_o=-641$, agree with   the  observed   thresholds.
The axial component $k_z$ of the wavevector 
decreases linearly with  $\epsilon$ from  a critical  value  $k^c_z$
close  to 0.30  (see fig.~\ref{fig:curves}(b)). As long as  the spiral
is well defined, the azimuthal wave  number $n_\theta$ equals 6  along
paths A and  B and 5 along path C. $n_\theta$ being  constant  and
$k_z$  decreasing with $\epsilon$,  the spiral  pitch, $\phi=\arctan
(k_\theta  / k_z)$, increases with  $\epsilon$ from a critical  value
$\phi^c=18^\circ$ up to $40^\circ$.

\begin{figure}
\vspace{-0.5cm}
  \begin{center}
    \psfig{figure=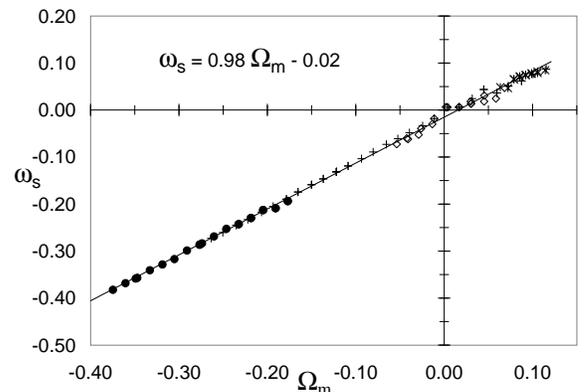, width=8cm, angle=0}
  \end{center}
\vspace{-0.5cm}
    \caption{The angular velocity $\omega_s$ vs.
$\Omega_m$, ($R_o=-641$ ($\ast$), $R_o=-713$ ($\diamond$),
$R_o=-900$ ($\bullet$) and $R_i=704$ ($+$)).} 
    \label{fig:omega}
\vspace{-0.0cm}
\end{figure}

We now consider the horizontal  path D with $R_i=704$. For $R_o>-630$,
the  flow  is fully  turbulent.  Decreasing  $R_o$, spiral  turbulence
emerges from  the fully turbulent  regime. The initial  azimuthal wave
number   $n_\theta=5$  increases   to  $n_\theta=6$   by  Eckhaus-like
re-arrangements.  The axial wave  vector component remains constant on
the  explored values  $-950<R_o<-630$.   Finally, fig.~\ref{fig:omega}
displays  $\omega_s=\omega/n_\theta$,  the  angular  velocity  of  the
spiral               turbulence              pattern,              vs.
$\Omega_m=\frac{1}{2}(\Omega_i+\Omega_o)$.  As seen  by Coles  and Van
Atta~\cite{Coles,VanAtta,Coles-VA}, the pattern  rotates with the mean
angular velocity of the two  cylinders ($\omega_s = \Omega_m$) and not
with   the  outside  cylinder   as  reported   by  Andereck   {\it  et
al.}~\cite{Andereck}     as    well     as    by     Goharadzeh    and
Mutabazi~\cite{GohMut}. In  the latter  study, the range  of variation
for  $\Omega_i/\Omega_o$  is  actually  too  small  to  distinguish  a
dependence on $\Omega_o$ from one on $\Omega_m$.

Altogether,  the spiral turbulence  emerges as  a wavy turbulent state
pattern, which bifurcates continuously  from the homogeneous turbulent flow. 
This wavy behavior agrees with the idea of a phase
dynamics analysis  suggested by Hegseth  {\it et al.}~\cite{Heg1,Heg2}
to describe the pitch variation, they observed. 
However the present study does not report any pitch variation. 
In their analysis, Hegseth {\it et al.}, on the basis of
the  mean  flow measurements  by  Coles and  Van~Atta~\cite{Coles-VA},
suggest  that the  spiral turbulence  induces a  backflow,  whose mean
axial  component is  differently  modified at  both extremities.   The
resulting different pitches locally imposed in the boundary conditions
drive the pitch  variation along the axis. In  the present study, both
the large  aspect ratio and  the gap thinness considerably  reduce the
backflow and the influence of the end boundaries.

For low  values of $R_i$,  we observed disconnected  turbulent domains
surrounded by  a laminar flow, which evokes  the transitional dynamics
observed   in  the   plane  Couette   flow~\cite{BotDavManDau},  where
turbulent domains  move, grow, decay,  split and merge leading  to the
so-called  spatio-temporal  intermittency (STI),  a  process in  which
active/turbulent regions  may invade absorbing/laminar  domains out of
which             disorder             cannot            spontaneously
emerged~\cite{ChatMan}. Considering the  vicinity of the plane Couette
flow     limit,     $\eta=1$,     and    following     Colovas     and
Andereck~\cite{ColAnd}, one would  like to know whether Taylor-Couette
flow actually exhibits STI and  how the flow evolves from a continuous
turbulent   state   modulated  in   space   and   time  to   two-state
coexistence.  In  a  first  attempt  to  answer  these  questions,  we
binarized the  flow into alternating laminar and  turbulent domains as
is common  in an STI  context.  This first-order  approximation should
not influence the  results presented below, at least  at a qualitative
level. 

\begin{figure}
\vspace{-0.5cm}
  \begin{center}
    \psfig{figure=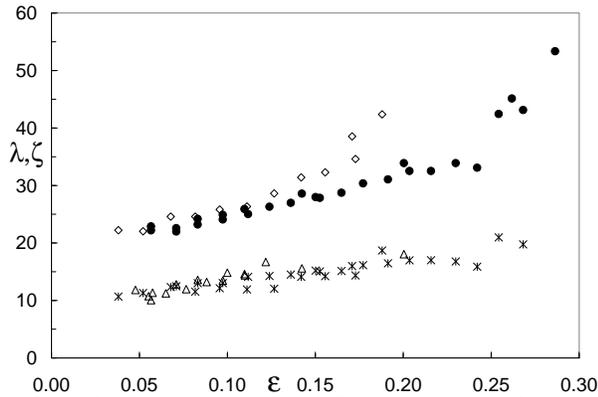, width=8cm, angle=0}
  \end{center}
\vspace{-0.5cm}
    \caption{$\lambda$ ($R_o=-713$ ($\diamond$) and $R_o=-922$
($\bullet$)), the pattern wavelength and $\zeta$ ($R_o=-713$
($\triangle$) and $R_o=-922$ ($\ast$)), the
turbulent stripes width, against $\epsilon$.} 
    \label{fig:lp}
\vspace{-0.3cm}
\end{figure}

\noindent
Fig.~\ref{fig:lp}   displays  the  turbulent   stripes  width,
$\zeta$, and the wavelength, $\lambda$ of the spiral pattern, parallel
to the wave vector direction. $\lambda$ increases faster than $\zeta$:
the minima  of the modulation becomes  more and more  distant from the
most active  turbulent regions. Simultaneously the  modulation amplitude
increases and these minima are less and less turbulent.  As a result they
become areas favorable to local relaxation to the laminar flow and
the  modulated  turbulent  state  turns  into a  periodic  pattern  of
alternating turbulent  and laminar  stripes separated by  fronts. This
totally  different  spatio-temporal dynamics  rapidly  turns into  the
disordered STI regime.

Our very large aspect ratio  TC apparatus has allowed, for the first
time, to visualize and quantitatively analyse the \lq\lq barber pole
turbulence\rq\rq at large scales.  First, this study brings quantitative
evidence of the spatio-temporal periodicity of the pattern and 
its dependance on control parameters.  Second, it
shows that the spiral  turbulence, previously described as alternating
turbulent and laminar bands,  actually appears through a modulation of
the homogeneous  turbulent state.  To our knowledge,  it is  the first
time  that  the  transition  from  homogeneous  turbulence  to  spiral
turbulence  is  investigated. Finally  measurements  of the  turbulent
stripes width relatively to the  pattern wavelength have allowed us to
describe  the  relaxation  mechanism  followed between  the  modulated
turbulent state and the laminar-turbulent coexistence.

We thank H. Chat\'e, P. Manneville and I. Mutabazi for
fruitful discussions, and C. Gasquet and V. Padilla for technical
assistance.

\end{multicols}


\vspace{-0.5cm}
\begin{references}
\vspace{-1.7cm}

\bibitem{Andereck}
C. D. Andereck, S. S. Liu and H. L. Swinney, J. Fluid Mech. {\bf 164},
155 (1986).

\bibitem{Feyn}
R. P. Feynman, {\it Lecture Notes in Physics} (Addison-Wesley Reading,
MA, 1964), vol. 2.

\bibitem{Coles}
D. Coles, J. Fluid Mech., {\bf 21}, 385 (1965).

\bibitem{VanAtta}
C. Van Atta, J. Fluid Mech., {\bf 25}, 495 (1966).

\bibitem{Coles-VA}
D. Coles and C. W. Van Atta, Phys. of Fluids Supp., S120 (1967).

\bibitem{Heg1}
J. Hegseth, C. D. Andereck, F. Hayot and Y Pomeau, Phys. Rev. Letters,
{\bf 62}, 257 (1989).

\bibitem{Heg2}
J. Hegseth, C. D. Andereck, F. Hayot and Y. Pomeau, Eur. J. Mech. B, {\bf 10}, 221 (1991).

\bibitem{DauDav}
O. Dauchot and F. Daviaud, Phys. of Fluids, {\bf 7}, 335 (1995).

\bibitem{BottChat}
S. Bottin and H. Chat\'e, Eur. Phys. J. B, {\bf 6}, 143 (1198). 

\bibitem{TCvisu}
A. Prigent and O. Dauchot, submitted to Physics of Fluids (1999).

\bibitem{Snyder}
H. A. Snyder, The Physics of Fluids, {\bf 11}, 1599 (1968).

\bibitem{Croq}
V. Croquette and H. Williams, Physica D, {\bf 37}, 300 (1989).

\bibitem{GohMut}
A. Goharadzeh and I. Mutabazi, in {\it $2^{ième}$ Colloque sur le
Chaos temporel et le Chaos spatio-temporel}, Rouen (1998).

\bibitem{BotDavManDau}
S. Bottin, F. Daviaud, P. Manneville and O. Dauchot, Eur. Phys. Lett.,
{\bf 43}, 171 (1998).

\bibitem{ChatMan}
H. Chat\'e and P. Manneville, in {\it Turbulence, A tentative dictionary},
P. Tabeling and O. Cardoso Eds., New York, p 111-116, Plenum Press (1994).

\bibitem{ColAnd}
P. W. Colovas and C. D. Andereck, Phys. Rev. E, {\bf 55}, 2736 (1997).
\end{references}
\end{document}